\documentstyle[preprint]{jpsj}
\catcode`\@=11
\newcount\@arga
\newcount\@argb
\newcount\@argc
\newcount\@ctmpa
\newcount\@ctmpb
\newcount\@ctmpc
\newcount\@ctmpd
\newcount\@ctmpe
\newdimen\@darg
\newdimen\@bblen
\newif\if@bbllx
\newif\if@bblly
\newif\if@bburx
\newif\if@bbury
\newif\if@height
\newif\if@width
\newif\if@scale
\newif\ifno@bb
\newif\ifepsfdraft
\epsfdraftfalse
\def\@setpsfile#1{
                \typeout{epsf:[#1]}
                \def\@psfile{#1}
}
\def\@setpsheight#1{
                \@heighttrue
                \@darg=#1\relax
                \edef\@psheight{\number\@darg}
}
\def\@setpswidth#1{
                \@widthtrue
                \@darg=#1\relax
                \edef\@pswidth{\number\@darg}
}
\def\@setpsscale#1{
                \@scaletrue
                \def\@pshscale{#1}
                \def\@psvscale{#1}
                \@bblen=#1pt\relax
                \@bblen=1000\@bblen
                \def\@texhscale{\expandafter\remove@dim\the\@bblen}
                \let\@texvscale=\@texhscale
}
\def\@setpshscale#1{
                \@scaletrue
                \def\@pshscale{#1}
                \@bblen=#1pt\relax
                \@bblen=1000\@bblen
                \def\@texhscale{\expandafter\remove@dim\the\@bblen}
}

\def\@setpsvscale#1{
                \@scaletrue
                \def\@psvscale{#1}
                \@bblen=#1pt\relax
                \@bblen=1000\@bblen
                \def\@texvscale{\expandafter\remove@dim\the\@bblen}
}

\def\@setparms#1=#2,{\@nameuse{@setps#1}{#2}}
\def\ps@init@parms{
                \@heightfalse \@widthfalse
                \no@bbfalse
                \def\@psbbllx{}\def\@psbblly{}
                \def\@psbburx{}\def\@psbbury{}
                \def\@psheight{}\def\@pswidth{}
                \def\@pshscale{1}\def\@psvscale{1}
                \def\@texhscale{1000}\def\@texvscale{1000}
                \def\@psfile{}
                \def\@sc{}
}
\def\parse@ps@parms#1{
                \@for\@epsfile:=#1\do
                   {\expandafter\@setparms\@epsfile,}}
\newif\ifnot@eof
\newread\ps@stream
\def\bb@search{
        \openin\ps@stream=\@psfile
        \no@bbtrue
        \not@eoftrue
        \catcode`\%=12\relax
        \ifeof\ps@stream\typeout{epsf: File not found}\fi
        \loop
                \read\ps@stream to \line@in
                \global\toks200=\expandafter{\line@in}\relax
                \ifeof\ps@stream \not@eoffalse \fi
                \@bbtest{\toks200}\relax
                \if@bbmatch\not@eoffalse\expandafter\bb@cull\the\toks200\fi
        \ifnot@eof \repeat
        \catcode`\%=14
}       

\catcode`\%=12
\newif\if@bbmatch
\def\@bbtest#1{\expandafter\@a@\the#1
\long\def\@a@#1
        \ifx\@bbtest#2\@bbmatchfalse\else\@bbmatchtrue\fi}
\def\bb@cull 
        \@ifnextchar\space{\@latexbug}{\bb@extract}}
\def\bb@extract #1 #2 #3 #4 {
        \message{BoundingBox: (#1bp,#2bp)--(#3bp,#4bp)}
        \@darg=#1 bp\edef\@psbbllx{\number\@darg}
        \@darg=#2 bp\edef\@psbblly{\number\@darg}
        \@darg=#3 bp\edef\@psbburx{\number\@darg}
        \@darg=#4 bp\edef\@psbbury{\number\@darg}
        \no@bbfalse
}
\catcode`\%=14

\def\compute@bb{
                \bb@search
                \ifno@bb \typeout{epsf: No BoundingBox}
                \stop
                \else
                \@arga=\@psbburx
                \advance\@arga by -\@psbbllx
                \edef\@bbw{\number\@arga}
                \@arga=\@psbbury
                \advance\@arga by -\@psbblly
                \edef\@bbh{\number\@arga}
                \fi
}
\def\in@hundreds#1#2#3{\@argb=#2 \@argc=#3
                     \@ctmpa=\@argb     
                     \divide\@ctmpa by \@argc
                     \@ctmpb=\@ctmpa
                     \multiply\@ctmpb by \@argc
                     \advance\@argb by -\@ctmpb
                     \multiply\@argb by 10
                     \@ctmpb=\@argb     
                     \divide\@ctmpb by \@argc
                     \@ctmpc=\@ctmpb
                     \multiply\@ctmpc by \@argc
                     \advance\@argb by -\@ctmpc
                     \multiply\@argb by 10
                     \@ctmpc=\@argb     
                     \divide\@ctmpc by \@argc
                     \@arga=#1\@ctmpe=0
                     \@ctmpd=\@arga
                        \multiply\@ctmpd by \@ctmpa
                        \advance\@ctmpe by \@ctmpd
                     \@ctmpd=\@arga
                        \divide\@ctmpd by 10
                        \multiply\@ctmpd by \@ctmpb
                        \advance\@ctmpe by \@ctmpd
                     \@ctmpd=\@arga
                        \divide\@ctmpd by 100
                        \multiply\@ctmpd by \@ctmpc
                        \advance\@ctmpe by \@ctmpd
                     \edef\@result{\number\@ctmpe}
}
\def\compute@wfromh{
                \in@hundreds{\@psheight}{\@bbw}{\@bbh}
                \edef\@pswidth{\@result}
}
\def\compute@hfromw{
                \in@hundreds{\@pswidth}{\@bbh}{\@bbw}
                \edef\@psheight{\@result}
}
\def\compute@handw{
        \if@height 
                \if@width
                \else
                        \compute@wfromh
                \fi
        \else 
                \if@width
                        \compute@hfromw
                \else
                        \if@scale
                                \in@hundreds{\@texvscale}{\@bbh}{1000}
                                \let\@bbh=\@result
                                \in@hundreds{\@texhscale}{\@bbw}{1000}
                                \let\@bbw=\@result
                        \fi
                                \edef\@psheight{\@bbh}
                                \edef\@pswidth{\@bbw}
                \fi
        \fi
}
{\catcode`\p=12\catcode`\t=12
\gdef\remove@dim#1.#2pt{#1}}

\def\compute@sizes{
        \compute@bb
        \compute@handw
}
\def\epsfile#1{
        \ps@init@parms
        \parse@ps@parms{#1}
        \compute@sizes
        \@arga=\@psheight
        \divide\@arga by 65536
        \edef\@psvsize{\number\@arga}
        \@arga=\@pswidth
        \divide\@arga by 65536
        \edef\@pshsize{\number\@arga}
        \message{=>(\@pshsize bp,\@psvsize bp)}
        \leavevmode
        \vbox to \@psheight true sp{
                \hbox to \@pswidth true sp{
                \ifepsfdraft\hss\@psfile\hss\else
                \if@height 
                        \if@width
                                \special{epsfile=\@psfile \space 
                                hsize=\@pshsize \space
                                vsize=\@psvsize \space}
                        \else
                                \special{epsfile=\@psfile \space 
                                vsize=\@psvsize \space}
                        \fi
                \else 
                        \if@width
                                \special{epsfile=\@psfile \space 
                                hsize=\@pshsize \space}
                        \else
                                \if@scale
                                        \special{epsfile=\@psfile \space
                                        vscale=\@psvscale \space
                                        hscale=\@pshscale \space}
                                \else
                                        \special{epsfile=\@psfile \space}
                                \fi
                        \fi
                \fi
                \hfil\fi
                }
        \vfil
        }
}

\catcode`\@=12

\def\runtitle{Impurity Effect on Spin Ladder System}
\def\runauthor{Yukitoshi {\sc Motome}, Nobuyuki {\sc Katoh},
Nobuo {\sc Furukawa} and Masatoshi {\sc Imada}}

\def\sf{\rm}

\title{Impurity Effect on Spin Ladder System}
\author{Yukitoshi {\sc Motome}, Nobuyuki {\sc Katoh},
Nobuo {\sc Furukawa} and Masatoshi {\sc Imada}
}
\inst{
Institute for Solid State Physics, University of Tokyo, \\
Roppongi 7-22-1, Minato-ku, Tokyo 106
}
\recdate{
\today
}

\abst{
Effects of nonmagnetic impurity doping in a spin ladder system 
with a spin gap are investigated by the exact diagonalization as well as
by the variational Monte Carlo calculations.
Substantial changes in macroscopic properties such as enhancements 
in spin correlations and magnetic susceptibilities 
are observed in the low impurity concentration region,
which are caused by the increase of low-energy states. 
These results suggest
that small but finite amount of nonmagnetic impurity doping
relevantly causes the reduction or the vanishment of the spin gap.
This qualitatively explains the experimental result 
of Zn-doped SrCu$_{2}$O$_{3}$ 
where small doping induces gapless nature.
We propose a possible scenario for this drastic change
as a quantum phase transition
in a spin gapped ladder system due to spinon doping effects.
}

\kword{
spin ladder model, spin gap, SrCu$_{2}$O$_{3}$, Zn doping,
nonmagnetic impurity, spinon doping, long-range RVB
}

\begin{document}
\sloppy
\maketitle

\def\NENP{Ni(C$_2$H$_8$N$_2$)$_2$NO$_2$(ClO$_4$)}
\catcode`\@=11
\def\simle{\mathrel{\mathpalette\@versim<}}   
\def\simge{\mathrel{\mathpalette\@versim>}}   
\def\@versim#1#2{\lower2.5pt\vbox{\baselineskip0pt \lineskip-.5pt
   \ialign{$\m@th#1\hfil##\hfil$\crcr#2\crcr\sim\crcr}}}
\catcode`\@=12

Quantum effects in low-dimensional antiferromagnetic
 spin systems have intensively attracted 
theoretical and experimental interests.
One of such effects is observed as a formation of a  spin gap,
which competes with  antiferromagnetic long-range order
instabilities.
The origin of the spin gap  arising from the
structural feature has been extensively studied in 
a unified way for various systems.\cite{NKatoh-JPSJ64-1437}

An example for such systems
is the integer spin antiferromagnetic Heisenberg (AFH) chain which exhibits the
Haldane gap.\cite{Haldane83}
For a finite size $S=1$ AFH chain with
an open boundary condition, 
 the existence of $S=1/2$ states at the
 ends of the chain has been shown
from the valence-bond solid (VBS) picture.\cite{Affleck87}
This situation is realized if we dope impurities which
disconnect the topological connection of the chain.
Provided that the impurity concentration is low,
nearly-free spin degrees of freedom localized around impurities
are activated in the low-temperature region while
the bulk spin gap structure remains almost unchanged.
 Experimentally, a typical $S=1$  quasi one-dimensional antiferromagnet
{\NENP} (NENP)  exhibits a spin gap behavior.\cite{Renard90}
The spin degrees of freedom for edge
states  have  been  observed in Cu-doped
 NENP,\cite{Hagiwara90} which gives an experimental evidence to
support the existence of the VBS state. 
In this material, impurity effect has played an important role
as a probe to investigate the microscopic properties.

In contrast with the $S=1$ chain,
recent experiments on spin-gapped system CuGeO$_{3}$
with spin-Peierls distortion have shown that
small amount of Zn doping induces the AF long-range order.
\cite{Oseroff,Hase}
Here, Zn$^{2+}$ ion
plays a role of nonmagnetic impurity on the parent Cu$^{2+}$
($S=1/2$) lattice.
To understand the origin of this contrast,
it seems to be important to consider
relatively large interchain coupling for CuGeO$_{3}$.

Another compound which shows the spin gap behavior
is the 2-leg spin ladder system, SrCu$_{2}$O$_{3}$.
\cite{ZHiroi-JSSC95-230,Azuma1} 
The origin is understood from
the short-range resonating valence bond (RVB) picture.\cite{Dagotto,White}
Recently, experimental studies of Zn-doped 
compounds Sr(Cu$_{1-x}$Zn$_{x}$)$_{2}$O$_{3}$
has been performed.\cite{Nohara,Azuma}
Experimental results suggest that, at $x \simge 0.01$, there exists
the antiferromagnetic phase in the low temperature region. Above the
critical temperature, thermodynamical properties are
similar to those of a quasi one-dimensional gapless antiferromagnet.
The phenomena observed in Sr(Cu$_{1-x}$Zn$_{x}$)$_{2}$O$_{3}$
 suggest that, unlike the case for impurity doping in NENP,
a substantial change has occurred in the bulk spin state
 by Zn doping of less than {1\%}.

In this paper, we investigate nonmagnetic impurity effects 
in a 2-leg spin ladder system at low concentration of impurities,   
which may correspond to the Zn doping effects in experiments. 
In the present results, the exact diagonalization (ED) of clusters as well 
as variational Monte Carlo (VMC) calculations show that 
even small concentration of nonmagnetic impurities reduces the spin gap 
and remarkably enhances the antiferromagnetic correlation.

The Hamiltonian is a conventional $S=1/2$ AFH
model with the nearest neighbor interaction written as 
${\cal H}=J \sum_{<i,j>} \mbox{\boldmath $S$}_{i}
\mbox{\boldmath $ \cdot S$}_{j}$ 
on the 2-leg ladder structure 
with the periodic boundary condition,
where $J$ represents the spin exchange coupling.
In the present work,
the nonmagnetic impurity effect is treated as
the annihilation of impurity sites in the parent system.
Since we are interested in impurity effects 
at low concentration of impurities, 
we focus on the case with two impurity sites which are apart from each 
other as far as possible, that is, two impurities are on the site 1 
and $3L/2+1$ in the $L \times 2$ system.
 (We assign the site number from 1 to $L$ on the first leg and 
from $L+1$ to $2L$ on the second leg. 
For example, in the $12 \times 2$ system, 
the two impurities are located at the site 1 and 19,
as shown in the inset of Fig.~\ref{fig:1}.)
The ground state becomes singlet when $L=4n$, where $n$ is an integer. 

We first show the spin correlations in the real space 
calculated by the ED method in Fig.~\ref{fig:1}. 
To investigate the impurity effect on the spin correlation
around the impurity site, we measure the spin correlation  
$(-1)^{|i-j|}
\langle \mbox{\boldmath $S$}_i \mbox{\boldmath $\cdot S$}_j \rangle$ 
with $i$ fixed at the nearest-neighbor site to the impurity.
In the pure system, the spin correlation decays sharply
since this system has a spin gap in the thermodynamic limit. 
\cite{Dagotto}
In the impurity doped system, 
the antiferromagnetic correlation is enhanced. 

Next, we show the ED results for 
the temperature dependence of the uniform magnetic susceptibility $\chi$
per spin in Fig.~\ref{fig:2}. 
In the pure system, the susceptibility decays exponentially 
as the decrease of temperature due to the finite spin gap. 
On the contrary, 
the behavior of $\chi$ for the impurity doped system 
is apparently different from that for the pure system. 
As a reference, the $S=1/2$ AFH chain with the periodic 
boundary condition is also shown in Fig.~\ref{fig:2}, 
which is a typical spin gapless system in the thermodynamic limit. 
For $T/J \simge 0.15$, it is found that
the temperature dependence of $\chi$ for the impurity doped system
is rather similar to that for the AFH chain than the ladder case.
The sudden decay of $\chi$ appears at temperature below $T/J \sim 0.15$
for both the impurity doped ladder system and the AFH chain, 
which may be due to a finite size effect. 
The important point of these results is 
that the ground state property may change 
qualitatively due to the impurity doping at low concentration. 

To understand the microscopic origin of above behaviors, 
we calculate the low energy spectrum by the ED method, 
as shown in Fig.~\ref{fig:3}. 
In the pure ladder system, 
the low-lying states are very sparse. 
However, when two impurities are injected to the pure system, 
the number of low-lying excited states increases. 
The low energy excitation spectrum of the impurity doped system is similar to 
that of the AFH chain rather than that of the ladder system.

In order to study effects of impurities 
at low concentration region in a systematic way,
we treat larger size systems with the VMC technique.
We show that variational wave functions in our calculation have
excellent properties in the small size systems
compared with the ED results.
In the variational approaches,
we can get physical intuitions from trial wave functions.

We use the RVB type wave function,
originally introduced for 
the $S=1/2$ AFH model on a square lattice\cite{Liang},
which is defined by,
\begin{equation}
\label{RVB wave function}
|\Psi \rangle = \sum_{{i_{\alpha} \in A} \atop{j_{\beta} \in B}}
h \left( R_{i_{1} j_{1}} \right) \cdot \cdot \cdot
h \left( R_{i_{n} j_{n}} \right) \cdot
\left(i_{1} j_{1}\right) \cdot \cdot \cdot
\left(i_{n} j_{n}\right),
\end{equation}
where $h(R_{ij})$ is a weight function
for a singlet pair $(ij)$ lying between the different sublattice $A$ and $B$,
whose form is determined variationally. 
Here, $R_{ij}$ is the Manhattan distance between the site $i$ and $j$.
In the impurity doped case,
we make the following ansatz;
$R_{ij}$ is defined as the length of the shortest path from the site $i$ to $j$
without passing over impurity sites.
In the present work,
we take two types of $h(R)$ with two variational parameters for each type.
Both have one free parameter $h(3)$
(we set the normalization as $h(1) = 1$)
because the energy is sensitive to the functional form of $h(R)$
in the short distance.
For $R \geq 5$, one type has an exponential tail;
$h(R) = h(3) \exp \left[ \kappa \cdot \left(3-R\right) \right]$,
and the other has an algebraic tail;
$h(R) = h(3) \left( 3/R \right)^{p}$.
Here, $\kappa$ and $p$ are variational parameters
controlling the tail of the weight function.
Among these wave functions,
the optimized state is chosen to give the lowest variational energy 
$\langle \Psi| {\cal H} | \Psi \rangle / \langle \Psi| \Psi \rangle$.
We focus on the same configuration of two impurities
as in the ED calculations. 
We also calculate for some other configurations of two impurities and
have qualitatively same results.

First, we check the efficiency of VMC
by comparing with the exact results for small size systems.
Our variational wave function (\ref{RVB wave function}) is found to give
good agreement with the exact ground state energy.
For example, for the $12\times2$ system,
variational energy per site are
$-0.5777\pm0.0001J$ for the pure case and
$-0.4890\pm0.0002J$ for the impurity doped case,
while the exact results are
$-0.5784J$ and $-0.4912J$ respectively.
The spin correlations for the $12\times2$ system
calculated with the optimized wave functions are shown in Fig.~\ref{fig:1}.
The apparent difference between the pure and impurity doped case 
is well reproduced within this VMC calculation.

We calculate larger-size systems up to $96\times 2$ sites.
For all the sizes, the variational wave function changes 
drastically by the impurity doping
in spite of very low concentration:
For pure cases, the wave function 
with the exponential decay form of $h(R)$ gives
lower energy than that with the algebraic decay form.
However, if impurities are introduced,
the latter has the lower energy.
(In the pure case, the energy difference between these two types is small,
for example, $\simeq 0.0003J$ for the $48\times2$ case
which is comparable to errorbars
due to the strong decay of $h(R)$ and the finite size effects.
On the other hand, in the impurity doped case,
the difference is considerable,
for example, $\simeq 0.0013J$ for the $48\times2$ system.)
The spin correlations 
calculated with the optimized wave functions
are shown in Fig.~\ref{VMCSiSj}.
For all the sizes,
the spin correlations decay exponentially in the pure case,
which indicates the existence of the spin gap by the short-range RVB.
Our results show that
impurities cause the remarkable enhancement of the spin correlations.

All above results suggest that, in the ladder system,
nonmagnetic impurities even at the low concentration
cause drastic changes:
As shown in the ED study,
the susceptibility $\chi$ and the low energy spectrum
in the impurity doped cases are apparently different
from those of the pure ladder system and
rather similar to those of the AFH chain which is a gapless system.
The variational wave function for the impurity doped cases
is optimized with the long-range RVB weights
which directly result in the remarkable enhancement of the spin correlation.
All these results indicate that the doped impurities
intensively reduce or destroy the spin gap.

We now discuss the effect of nonmagnetic impurity doping
on the ladder system.
The undoped state, that is,
the pure ladder system has a finite spin gap $\sim 0.5J$.\cite{Dagotto}
This state can be qualitatively understood 
as the resonating state of dimer singlets,\cite{White}
schematically depicted in Fig.~\ref{picture}(a).
The special topological character of the 2-leg ladder favors
this dimer gapped state,
rather than the long-range RVB state
which generally contains more singlet states resonating to each other
in the summation of eq.~(\ref{RVB wave function}).
A doped impurity breaks a dimer singlet pair
and leaves an unpaired spin,
as shown in Fig.~\ref{picture}(b).
In this sense, the nonmagnetic impurity doping can be considered
as the doping of topological solitons with $S=1/2$
or so-called spinons.
\cite{Kivelson}
So far as the impurity concentration is very low
and randomly doped impurities are far apart from each other on average,
induced spinons may be localized near the impurity sites:
As depicted in Fig.~\ref{picture}(c),
the motion of spinons rearranges the dimer configuration.
Within this type of `staggered' dimer configuration
in the traces of moving spinons,
dimer singlets cannot resonate to each other.
Therefore, the system loses the resonance energy
proportional to distances between spinons and impurity sites,
which may cause the pinning of spinons.\cite{edge state}
In this phase, the spin correlation decays exponentially.

Next, we consider what happens
when the impurity concentration increases gradually.
In the ladder system,
the nonmagnetic impurity doping may not break
the one-dimensional structure at low impurity concentration.
\cite{S=1 system}
The increase of the number of spinons
nearly equal to the number of impurities
drives spinons to move more coherently in order to gain more kinetic energy.
The motion of spinons might 
destruct the short-range dimer structure and
totally reconstruct them to have longer bond lengths,
as schematically depicted in Fig.~\ref{picture}(d).
Consequently, the gains for both the spinon kinetic energy 
and the RVB resonance energy will be reconciled
by the RVB with the weight given by the algebraic decay.
Due to the power-law tail of RVB weights,
the bulk spin gap may be intensively reduced or destroyed
and the spin correlation may decay algebraically.
In this state, spinons are also involved
in the reconstructed long-range RVB,
as shown in Fig.~\ref{picture}(d).
Therefore, we cannot distinct spinons from other RVB any more, 
which means that the particle picture of spinons should break down.

The above transition,
characterized by the change of the spin correlation
from exponential decay to algebraic decay,
takes place 
when the impurity concentration increases beyond a critical value.
Our results for the finite size systems suggest that 
the phase with reconstructed long-range RVB
is realized even at a very small concentration of impurities
$\simge 0.01$ (as seen in $96\times2$ case).
It should be noted that the structural character of the ladder system
may have an important role to the change of the bulk properties. 
These phenomena can be considered as a new type of
the quantum phase transition induced by the spinon doping.

Our results may explain the experimental results
on Sr(Cu$_{1-x}$Zn$_{x}$)$_{2}$O$_{3}$ mentioned before.
The specific heat data above the critical temperature\cite{Nohara,Azuma} 
can be understood as the gapless nature due to the long-range RVB 
at the low doping region of nonmagnetic impurities.
Nevertheless, there still remain some problems
which seem to be beyond the framework of our present study.
For example, Sr(Cu$_{1-x}$Zn$_{x}$)$_{2}$O$_{3}$ shows
the AF long-range order at low temperature
and the $x$ dependence of the Curie constant 
slightly above the critical temperature.
\cite{Nohara,Azuma}
Some additional factors which exist in the real system,
such as the interladder coupling, the frustration in the spin exchange
or the random configurations of impurities,
may be relevant to these problems.
We leave them for further study. 

To summarize,
we have investigated the impurity effect on the spin gapped system.
The nonmagnetic impurities are doped into the 2-leg ladder system
and various physical properties are investigated
by the exact diagonalization and the variational Monte Carlo method.
Our results suggest that
the doping of nonmagnetic impurities into the ladder system
leads to drastic changes above very small critical concentration
with a reduction or disappearance of the bulk spin gap and
a remarkable enhancement of the spin correlation.
We propose a possible scenario for this substantial change.
These phenomena observed here can be considered as
a new type of the quantum phase transition induced by the spinon doping.
An experimental relevance of this spinon doping effects is
also mentioned.

The authors thank M. Azuma and M. Nohara
for stimulating discussions.
We have used a part of codes provided by H. Nishimori in TITPACK Ver.2.
This work is supported by a Grant-in-Aid for Scientific Research
on the Priority Area `Anomalous Metallic State near the Mott transition'
from the Ministry of Education, Science and Culture, Japan.
The computation in this work has been done
using the facilities of the Supercomputer Center,
Institute for Solid State Physics, University of Tokyo.

\newpage
\section*{Figure Captions}

\begin{figure}
\caption{
Spin correlations in the real space 
on the same leg for the $12\times2$ systems obtained by the ED calculation. 
Site index $i$ is fixed at site $2$ and 
index $j$ takes site $3,4,5,6,7$.   
(Inset shows the assignment of site number in these systems.)
Open symbols represent the spin correlations of 
the pure ladder system, 
while 
filled symbols are for 
those of the ladder system with two impurities. 
Square symbols are for the ED results, 
while circles show the VMC results for comparison.
}
\label{fig:1}
\end{figure}

\begin{figure}
\caption{
Temperature dependence of the uniform magnetic susceptibility per spin 
calculated by the ED method. 
Broken, solid and dash-dotted line correspond to the susceptibilities 
of the pure ladder system of the $6\times2$ size, the ladder system with 
two impurities of the $8\times2$ size and 
the $S=1/2$ AFH chain with 14 sites, respectively.
}
\label{fig:2}
\end{figure}

\begin{figure}
\caption{
Excitation spectrum in the low-energy region obtained by the ED method. 
Horizontal axis represents the excited energy 
scaled by the spin exchange $J$
which is measured from the ground state energy. 
Figures (a),(b) and (c) correspond to the low-energy spectrums of 
the pure ladder system of the $6\times2$ size, 
the ladder system with two impurities of the $8\times2$ size and 
the $S=1/2$ AFH chain with 14 sites, respectively.
}
\label{fig:3}
\end{figure}

\begin{figure}
\caption{Semi-log plot of spin correlations in the real space obtained by VMC.
We take the similar assignment for sites $i$ and $j$
as the ED study in Fig.~\ref{fig:1}.
For all sizes,
open and filled symbols show the pure and impurity doped cases, respectively.
}
\label{VMCSiSj}
\end{figure}

\begin{figure}
\caption{Physical pictures for
the pure and impurity doped ladder systems.
(a) A typical configuration of short-range RVB in the pure case.
The dotted RVB states show a resonance of singlet pairs.
(b), (c) and (d) Three different pictures on the doped state.
Gray ovals represent singlet pairs.
Crosses and arrows show impurity sites and spinons, respectively.
See text for details.}
\label{picture}
\end{figure}

\newpage
\setcounter{figure}{0}

\begin{figure}
\hfil
\epsfile{file=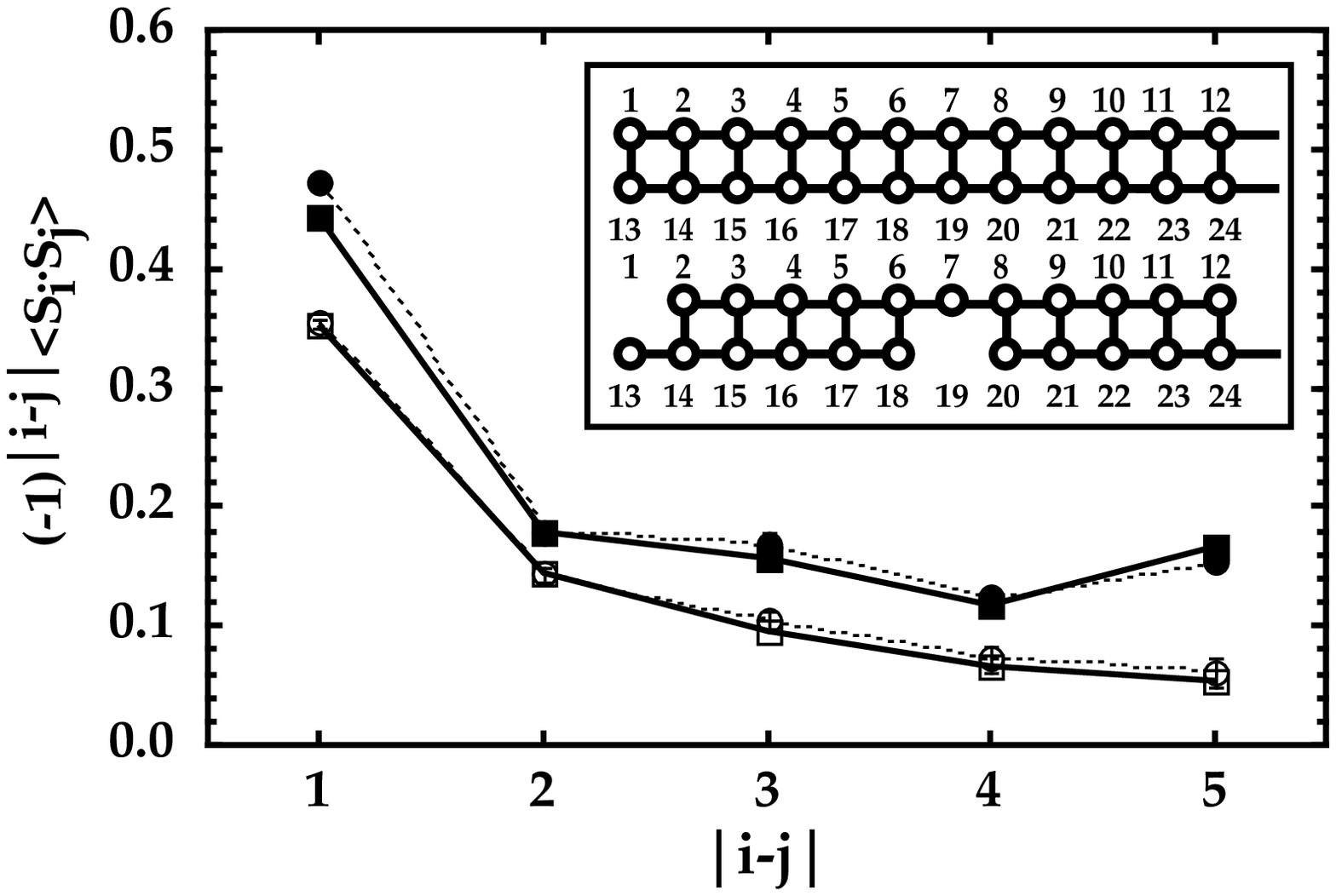,scale=1.0}
\hfil
\caption{
Y. Motome, N. Katoh, N. Furukawa and M. Imada
}
\end{figure}

\begin{figure}
\hfil
\epsfile{file=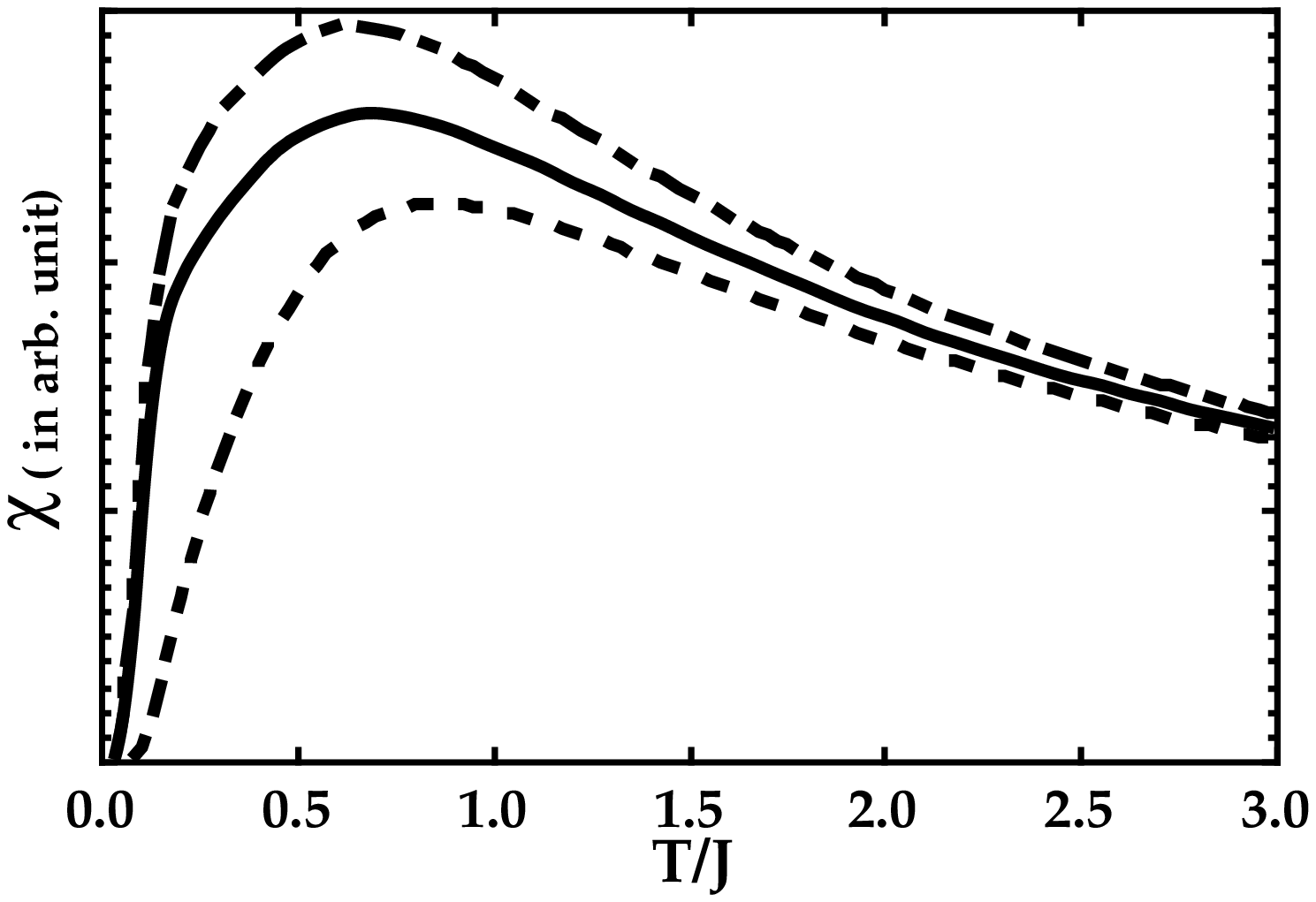,scale=1.0}
\hfil
\caption{
Y. Motome, N. Katoh, N. Furukawa and M. Imada
}
\end{figure}

\begin{figure}
\hfil
\epsfile{file=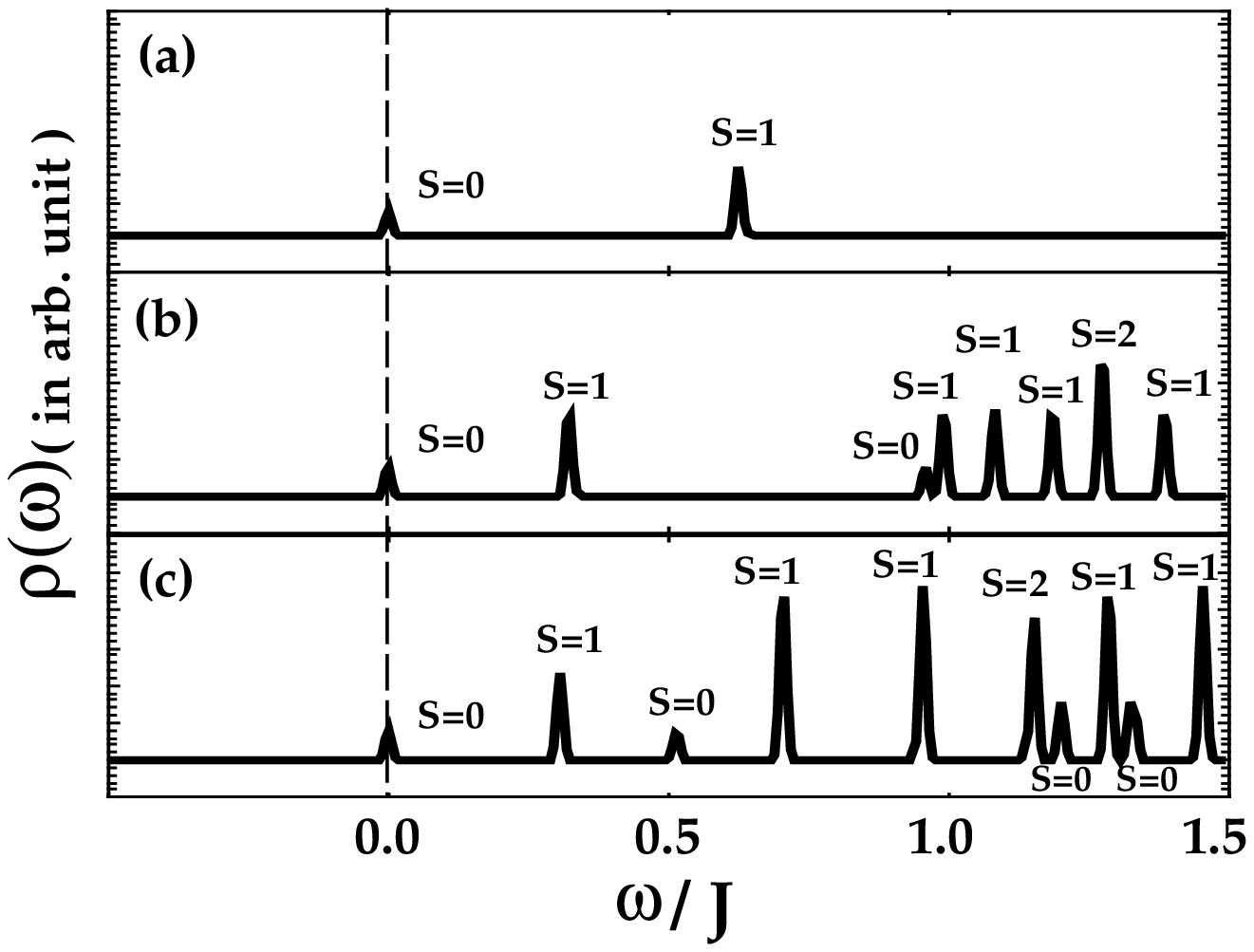,scale=1.0}
\hfil
\caption{
Y. Motome, N. Katoh, N. Furukawa and M. Imada
}
\end{figure}

\begin{figure}
\hfil
\epsfile{file=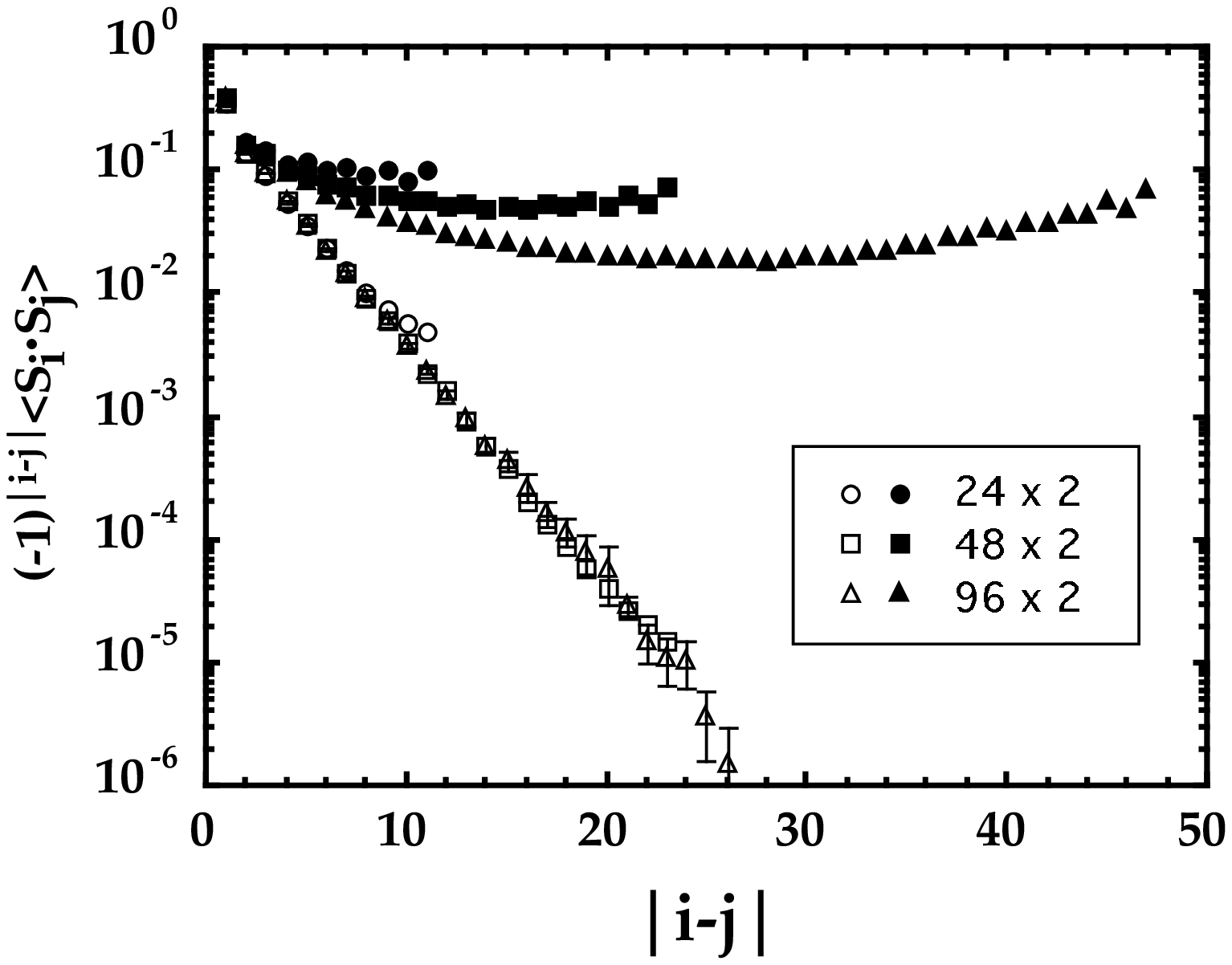,scale=1.0}
\hfil
\caption{
Y. Motome, N. Katoh, N. Furukawa and M. Imada
}
\end{figure}

\begin{figure}
\hfil
\epsfile{file=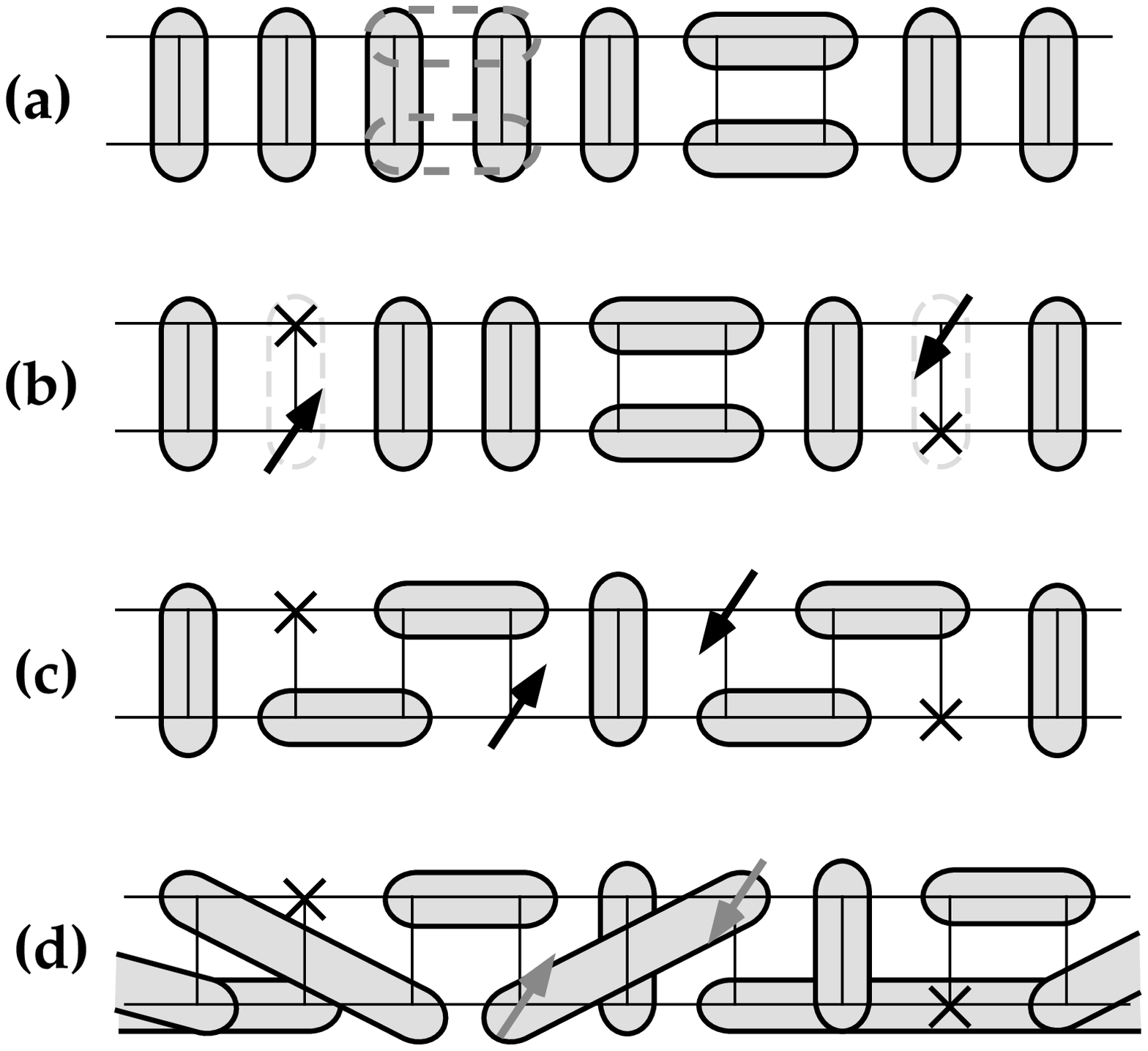,scale=1.0}
\hfil
\caption{
Y. Motome, N. Katoh, N. Furukawa and M. Imada
}
\end{figure}


\begin{thebibliography}{99}
\bibitem{edge state}
	This situation is related to the confinement
	of a local moment in the edge of the open ladder 
	with an unpaired edge site
	investigated in ref. \citen{White}.
\bibitem{S=1 system}
	In the low doping concentration region,
	impurities rarely disconnect the multi-leg ladder structure
	when they are doped randomly.
	The disconnection of the ladder is
	one of the randomness effects	
	which we neglect in this paper.
	On the other hand,
	the $S=1$ chain system is always disconnected by doped impurities.
	Therefore, induced $S=1/2$ spins at the ends of each open chain
	cannot obtain quantum coherency so that the bulk gap structure
	is not destroyed upon impurity doping.
\end{thebibliography}
\end{document}